\documentclass[10pt]{iopart}

\usepackage{color,graphicx}
\graphicspath{ {images/} }
\usepackage{iopams}
\expandafter\let\csname equation*\endcsname\relax
\expandafter\let\csname endequation*\endcsname\relax
\usepackage{amsmath}
\usepackage{hyperref}
\def\EAG{ErAl$_2$Ge$_2$}

\begin{document}

\title{Anisotropic magnetic properties of trigonal ${\rm ErAl_2Ge_2}$ single crystal}

\author{Moumita Nandi, A. Thamizhavel and S. K. Dhar}
\address{Department of Condensed Matter Physics and Materials Science, \\Tata Institute of Fundamental Research, Dr. Homi Bhabha Road, Colaba, Mumbai, 400 005, India}

%\ead{sudesh@tifr.res.in}



\begin{abstract}
We report the anisotropic magnetic properties of the ternary compound \EAG.\ Single crystals of this compound were grown by high temperature solution growth technique,using Al:Ge eutectic composition as flux.  From the powder x-ray diffraction  we confirmed that \EAG\ crystallizes in the trigonal CaAl$_2$Si$_2$-type crystal structure.  The anisotropic magnetic properties of a single crystal were investigated by measuring the magnetic susceptibility, magnetization, heat capacity and electrical resistivity.  A bulk magnetic ordering occurs around 4~K inferred from the magnetic susceptibility and the heat capacity. The magnetization measured along the $ab$-plane increases more rapidly than along the $c$-axis suggesting the basal $ab$-plane as the easy plane of magnetization. The magnetic susceptibility, magnetization and the $4f$-derived part of the heat capacity in the paramagnetic regime analysed based on the point charge model of the crystalline electric field  (CEF) indicate a relatively low CEF energy level splitting. 
\end{abstract}

%\begin{keyword}
%\EAG \sep magnetocrystalline anisotropy \sep single crystal \sep antiferromagnetism

%\end{keyword}

\maketitle
\ioptwocol

\section{Introduction}

The rare-earth intermetallic compounds of  the general formula $RT_2X_2$, where $R$ is the rare-earth, $T$ is a transition metal and $X$ is a $p$-block element, crystallizing in the ThCr$_2$Si$_2$-type crystal structure have been investigated  extensively owing to their interesting physical properties like superconductivity, heavy fermion behaviour, valence fluctuation, magnetic ordering etc.~\cite{steglich1979superconductivity, dung2009magnetic, thamizhavel2007anisotropic, joshi2010magnetocrystalline, drachuck2016magnetization}.  A lesser number of rare-earth based compounds with the general formula (1-2-2) crystallize in the trigonal  CaAl$_2$Si$_2$-type  crystal structure.   The reason for their relative paucity is that the number  of valence electrons should not exceed 16~\cite{klufers1984alpha, kranenberg2000structure}, which is satisfied by a lesser number of compositions.  However, it has been found that RAl$_2$X$_2$ (X = Si and Ge) compounds, where valence electron count is 17, are exceptions to this rule~\cite{kranenberg2000structure, kranenberg2000investigations}.  Kranenberg et al. attributed the stability of these  compounds to the small electronegativity differences between the Al and Si (Ge) atoms.    

The $R$Al$_2$X$_2$ compounds crystallizing in the  CaAl$_2$Si$_2$-type structure adopt the \textit{P\={3}m1} space group (\#164).  An interesting feature of this structure type is that the rare-earth atoms form a triangular lattice  in the $ab$-plane and the layers formed by Al and Si atoms are separated by distance $c$, the lattice parameter normal to the hexagonal plane.  The RAl$_2$X$_2$ (R = Eu and Yb; X = Si and Ge) have been previously investigated in the polycrystalline form~\cite{kranenberg2000structure, schobinger1989magnetic}.  We have reported the physical properties of  EuAl$_2$Si$_2$ single crystal~\cite{maurya2015anisotropic}, which orders antiferromagnetically at 33~K and exhibits a substantially large magnetoresistance (~1200~\% at 14~T) at 2~K for the field applied along the $c$-axis.    Very recently, we reported the magnetic properties of HoAl$_2$Ge$_2$ single crystal~\cite{matin2018single} which undergoes a bulk antiferromagnetic transition at 6.5~K with the $ab$-plane as the easy plane of magnetization.   In continuation of our studies on this series of compounds we have successfully grown the single crystals of ErAl$_2$Ge$_2$ and probed the magnetic behavior using the techniques of magnetization, electrical resistivity and heat capacity.

\section{Experimental methods}

Single crystals of \EAG\ were grown by the high temperature solution growth method,  taking advantage of the deep eutectic (420~$^{\circ}$C) formed by the Al:Ge (72:28)~\cite{Okamoto1993}.  High purity metals of Er, Al and Ge with a starting composition of  $1 : 17.5 :   7.5$ were placed in a high quality recrystallized alumina crucible.  The alumina crucible was sealed in an evacuated quartz ampoule under a partial pressure of argon gas.  The pressure of the argon gas was kept at a level such that it does not exceed the  atmospheric pressure at the maximum growth temperature.  The ampoule was placed in a resistive heating box type furnace and heated to 1050~$^{\circ}$C at a rate of 30~$^{\circ}$C/hr and held at this temperature for 20~hr for homogenizing the melt.  Then the furnace was cooled at the rate of 1.8~$^{\circ}$C/hr down to 600~$^{\circ}$C at which point the excess flux was removed by means of centrifuging.  Well defined shiny single crystals with typical  dimensions of 4~mm~$\times$~3~mm~$\times$~1~mm were obtained.    A few pieces of the single crystals were ground for powder x-ray diffraction measurement using PANalytical X-ray machine with a monochromatic Cu-K$_{\rm \alpha}$ radiation.  The magnetic measurements were performed using SQUID magnetometer (Quantum Design, USA) and the heat capacity and electrical measurements were performed using a physical property measurement system (PPMS).

\section{Results and Discussions}
%*********************FIGURE 1********************************
\begin{figure}[b]
	\includegraphics[width=0.5\textwidth]{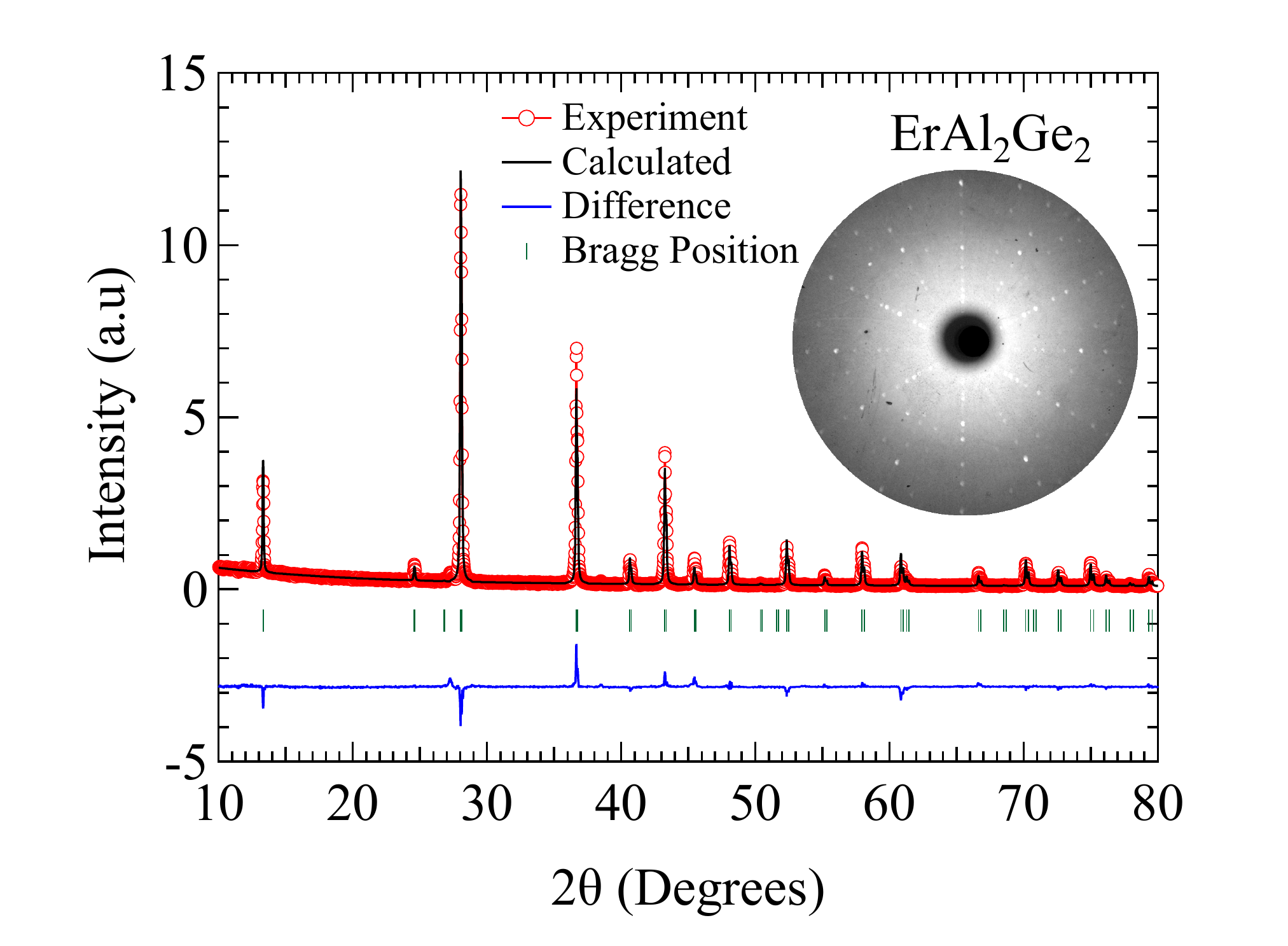}
	\caption{(Color online) Rietveld refinement of the powder x-ray diffraction pattern of crushed single crystals of \EAG.   Inset shows the Laue diffraction pattern corresponding to the (001) plane. }
	\label{Fig1}
\end{figure}
%****************************************************************
\subsection{X-ray studies}

The room temperature powder x-ray diffraction (XRD) pattern of \EAG\ is shown in Fig.~\ref{Fig1}.  All the peaks can be indexed to the trigonal CaAl$_2$Si$_2$ type crystal structure and there are no extra peaks due to any trace impurity phase(s).  The Rietveld analysis of the XRD data furnishes the  lattice constants as $a = 4.180$~\AA\ and $c~=~6.654$~\AA, which are in close agreement with the previously reported values~\cite{qin2008investigation}.    The well defined Laue diffraction pattern, shown in the inset of Fig.~\ref{Fig1} confirms the good quality of the single crystal.  The nearest Er-Er distance is in the $ab$-plane which is the same as that of the lattice constant $a$.

\subsection{Electrical resistivity}

%*********************FIGURE 2********************************
\begin{figure}[b]
	\includegraphics[width=0.5\textwidth]{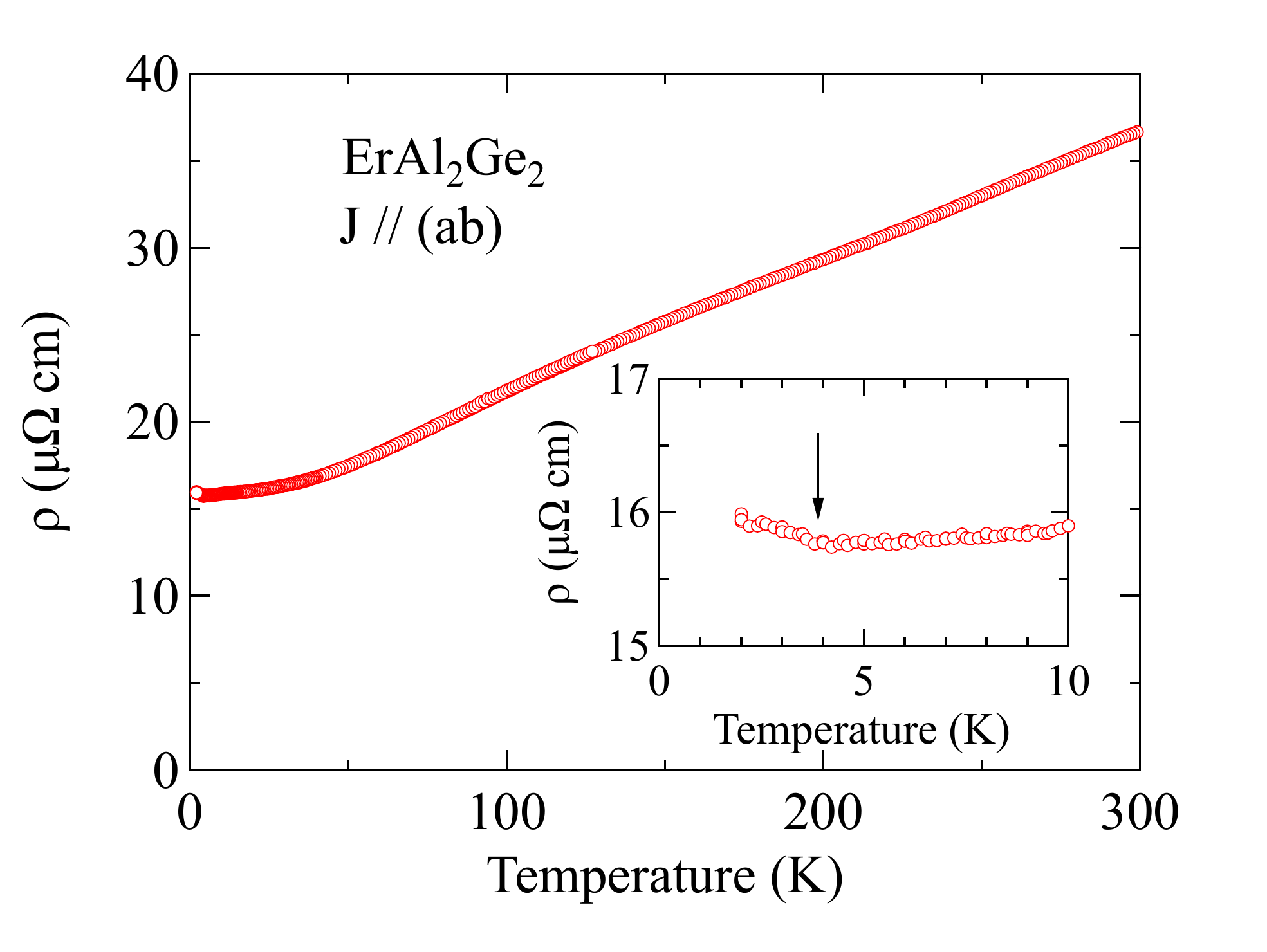}
	\caption{(Color online) Temperature dependence of electrical resistivity with current in the basal plane. The inset shows the low temperature region where the magnetic ordering is indicated by an arrow.  }
	\label{Fig2}
\end{figure}
%****************************************************************

The temperature dependence of electrical resistivity  from 2 to 300~K, with the current in the $ab$-plane, is shown in Fig.~\ref{Fig2}.  The electrical resistivity decreases with the decrease in temperature typical of a metal and shows an upturn below 4~K, which is due to the onset of antiferromagnetic ordering (see magnetization data below).  Similar  upturn in the electrical resistivity was observed in  the isostructural compound HoAl$_2$Ge$_2$~\cite{matin2018single}, below its N\'{e}el temperature.  The increase in the electrical resistivity in the antiferromagnetic state is usually attributed to the formation of superzone gap, when the magnetic periodicity is different from the lattice periodicity.  Because of the gap opening the number of charge carrier decreases which leads to the increase in the electrical resistivity.  The superzone gap is observed in elemental rare-earth metals  such as Dy, Er, Ho and Tm etc.~\cite{elliott1963theory} and has also been observed in several  rare-earth intermetallic compounds like CeGe, CePd$_5$Al$_2$, UCu$_2$Sn, UNiGa etc. to name a few~\cite{das2012anisotropic, onimaru2008giant, takabatake1998superzone, aoki1996superzone} .

\subsection{Magnetic susceptibility and isothermal magnetization}

%*********************FIGURE 3********************************
\begin{figure}[b]
\includegraphics[width=0.5\textwidth]{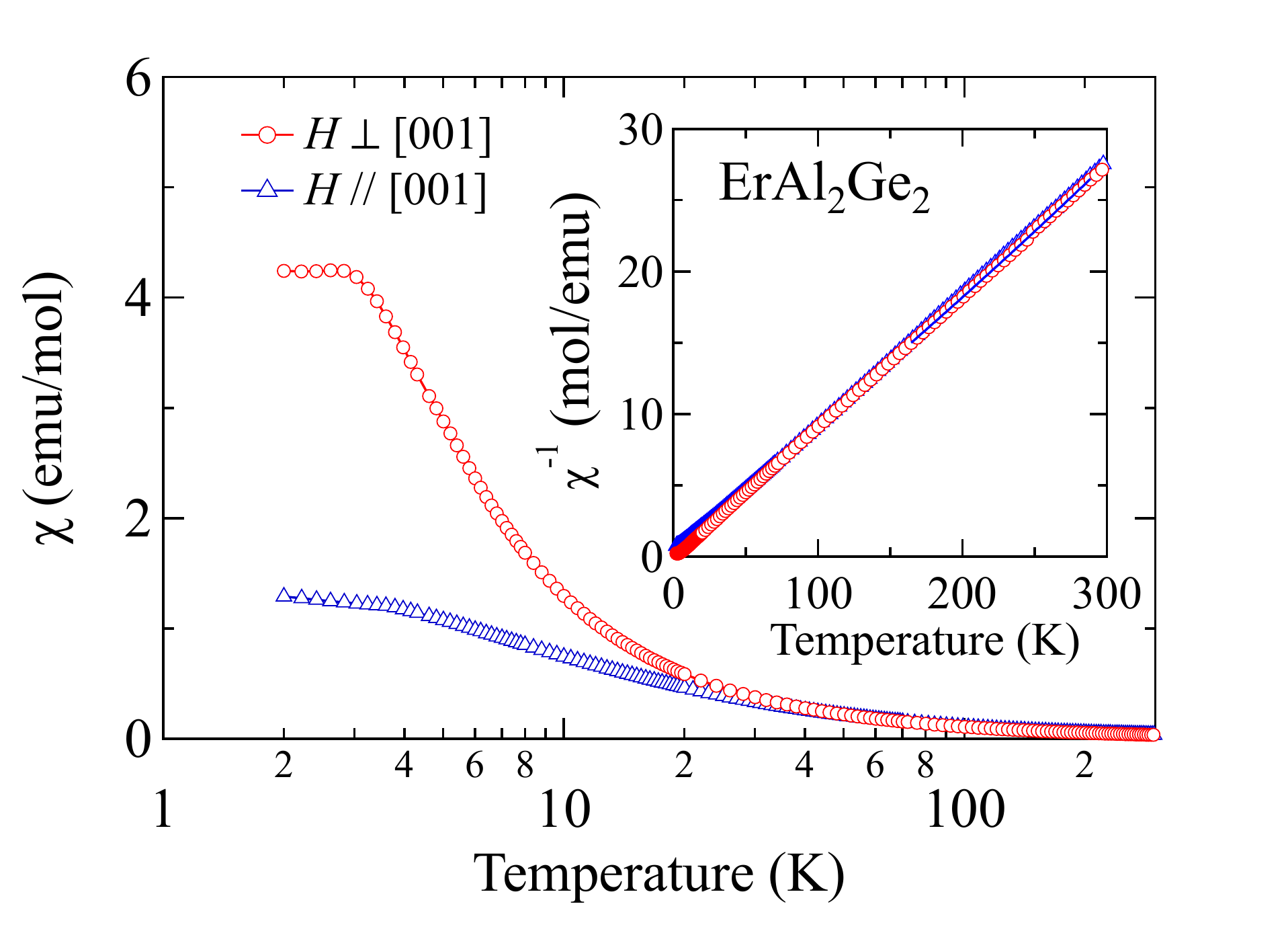}
\caption{(Color online) Temperature dependence of the magnetic susceptibility for field parallel and perpendicular to the $ab$-plane. The inset shows the inverse susceptibility plot along with the modified Curie-Weiss fit.  }
\label{Fig3}
\end{figure}
%****************************************************************

\paragraph{Magnetic susceptibility} The $dc$ magnetic susceptibility (main panel)  and the inverse magnetic susceptibility (inset)  of \EAG\ are shown in Fig.~\ref{Fig3}.  The susceptibility follows the Curie-Weiss behaviour at high temperatures.  There is a prominent change of slope in the susceptibility at around 4~K for field applied  both along the $c$-axis and normal to it, which is due to the antiferromagnetic transition at 4~K. A large anisotropy is observed at low temperatures (below 20~K) but it weakens considerably at higher temperature in the paramagnetic state.  The susceptibility in the $ab$-plane increases more rapidly  below 20~K  than along the $c$-axis, thus indicating that the easy axis of magnetization is in the $ab$-plane.  Similar behaviour was previously observed in HoAl$_2$Ge$_2$~\cite{matin2018single}. The inverse susceptibility data were fitted to the modified Curie-Weiss law, $\chi^{-1} =  (\chi_0 + \frac{C}{T-\theta_p})^{-1}$, where $\chi_0$ is the temperature independent term whose contributions come from the core-electrons and the Pauli spin susceptibility of the conduction electrons.  We obtain an effective magnetic moment of 9.73~$\mu_{\rm B}$/Er and 9.31~$\mu_{\rm B}$/Er and the paramagnetic Curie-Weiss temperature of $-7.35$~K and $-0.39$~K for $H~\parallel~ab$-plane and $c$-axis, respectively.  The  effective magnetic moment values are close to the Hund's rule derived value of 9.59~$\mu_{\rm B}$/Er for free Er$^{3+}$. If we compare the overall magnetic susceptibility with that of HoAl$_2$Ge$_2$,  the anisotropy in paramagnetic state is relatively weaker for ErAl$_2$Ge$_2$. This can be attributed to the crystal electric field (CEF) effect. A similar behavior is observed in R$_2$CoGa$_8$ (R = Gd-Lu), where the sign of the $B_2^0$ parameter (discussed later) changes as one moves towards higher rare-earth side~\cite{joshi2008anisotropic}.

\begin{figure}[b]
	\includegraphics[width=0.5\textwidth]{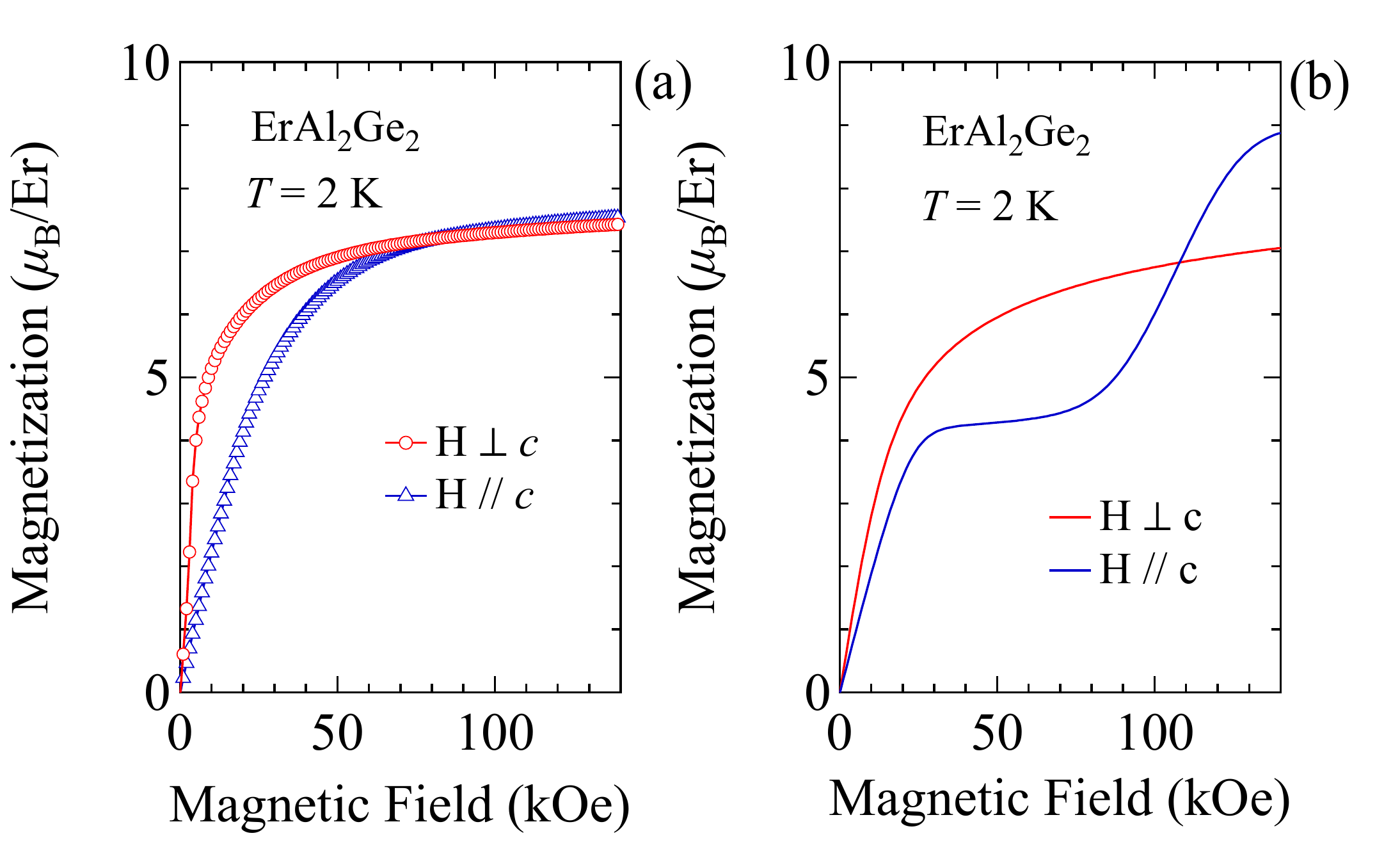}
	\caption{(Color online) (a) Isothermal magnetization at $T = 2$~K for $H~\parallel~(ab)$-plane and $H~\parallel~c$-axis for fields up to 14~T.  (b) Simulated magnetization plots with the CEF parameters (see text for details) . }
	\label{Fig4}
\end{figure}

\paragraph{Magnetization} The isothermal magnetization measured at $T= 2$~K is shown in Fig.~\ref{Fig4} for field parallel to $ab$-plane and $c$-axis, respectively.  The magnetization increases more rapidly in the $ab$-plane  at low fields and shows signs of gradual saturation as the field is increased above 10~kOe.   On the other hand, for  $H~\parallel~c$-axis the magnetization in comparison initially increases less rapidly at low fields and it is less than the corresponding value for $H~\parallel~ab$ up to a field of 85~kOe beyond which the magnetization along the $c$-axis crosses that of $ab$-plane thus indicating the change in easy axis at high magnetic fields.  It is interesting to mention here that the calculated magnetization based on a crystalline electric field model (to be discussed later) also exhibits a cross-over at slightly higher fields thus qualitatively matching with the experimental data as shown in Fig.~\ref{Fig4}(b).   The overall field dependence of magnetization behavior suggests an antiferromagnetic ordering in \EAG.   At 140~kOe, the magnetization attains a value of $7.43~\mu_{\rm B}$/Er  and  $7.54~\mu_{\rm B}$/Er, respectively in the $ab$-plane and along the $c$-axis, respectively.  These values are lower than the saturation moment of Er$^{3+}$ which is given by $g_{\rm J}J (\frac{6}{5} \times  \frac{15}{2}=)$ 9~$\mu_{\rm B}$/Er.   Apparently higher fields are necessary to attain the full moment of Er$^{3+}$.

%*********************FIGURE 4********************************
%****************************************************************

\paragraph{Crystal electric field analysis}  We have applied  the point charge model of  crystalline electric field to the magnetic susceptibility data to get a semi-quantitative estimate of the CEF level splitting.  The Er-atom in \EAG\ unit cell, occupies the $1a$ Wyckoff's position with  point symmetry $\bar{3}m$ (Sch\"{o}nflies symbol: $D_{3d})$ which possesses trigonal site symmetry.  For the sake of convenience, we  have used the CEF Hamiltonian for the hexagonal site symmetry which is given by,

\begin{equation}
\label{eqn4}
\mathcal{H}_{\rm CEF} = B_2^0{\bf O}_2^0 + B_4^0{\bf O}_4^0 + B_6^0{\bf O}_6^0 + B_6^6{\bf O}_6^6,
\end{equation}

where $B_{\rm n}^{\rm m}$ are the crystal field  parameters and ${\bf O}_{\rm n}^{\rm m}$ are the Stevens operators~\cite{stevens1952matrix, hutchings1965solid}.   The 16-fold $(2J +1; J = 15/2)$ degenerate level of free Er$^{3+}$ ion splits into 8 doublets in the CEF of hexagonal symmetry.  The magnetic susceptibility including the molecular field contribution $\lambda$ is given by

\begin{equation}
\label{eqn5}
\chi^{-1} = \chi_{\rm CEF}^{-1} - \lambda_i,
\end{equation}

%*********************FIGURE 5********************************
\begin{figure}[b]
\includegraphics[width=0.5\textwidth]{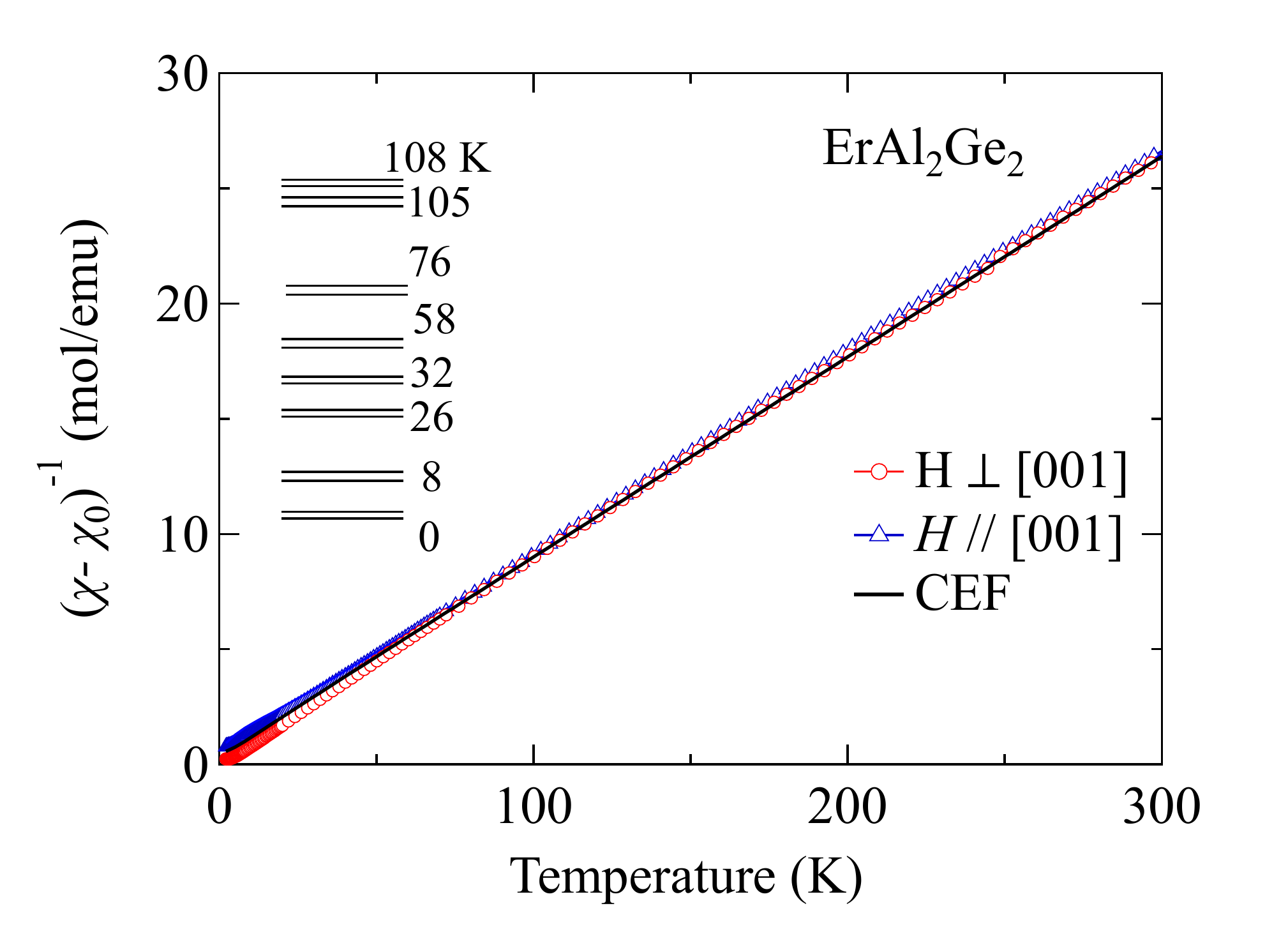}
\caption{(Color online) The crystal electric field analysis on the inverse susceptibility data.  The solid lines are fit to the CEF analysis (refer to text).  The estimated energy levels are also shown. }
\label{Fig5}
\end{figure}
%****************************************************************

where $\chi_{\rm CEF}$ is CEF susceptibility.  The expression for the magnetic susceptibility and magnetization based on the CEF model is given in Ref.~\cite{das2011magnetic, das2014anisotropic}.  In order to analyse the inverse susceptibility in the above CEF model, we first  fitted the paramagnetic inverse susceptibility to modified Curie-Weiss expression by fixing the $\mu_{\rm eff}$ value to 9.59~$\mu_{\rm B }$/Er and subtracted the resulting $\chi_0$ value from the raw susceptibility data.  Finally we plotted the inverse $(\chi - \chi_0)$ versus temperature and fitted  the CEF Hamiltonian of Eqn~\ref{eqn5} to the renormalized data, following the procedure applied earlier on some rare-earth intermetallic compounds~\cite{takeuchi2004magnetism, mondal2018magnetocrystalline}.  The solid lines in Fig~\ref{Fig5} show the fitted inverse magnetic susceptibility based on the CEF model, with the crystal field parameters as $B_2^0 = -0.062$~K, $B_4^0 = 0.001$~K, $B_6^0 = -3.9~\times~10^{-5}$~K, $B_6^6 = 2.0~\times~10^{-5}$~K.  The molecular field coefficients are  $\lambda_x = -0.1$~mol/emu and $\lambda_z = -0.62$~mol/emu.  The  negative sign of the molecular field constants, albeit small, supports the  antiferromagnetic nature of the magnetic ordering.  The corresponding energy eigenvalues of the crystal field levels are also shown in Fig.~\ref{Fig5}.  In practice various combinations of CEF parameters $B_n^m$'s which furnish widely different CEF energy eigenvalues provide a good fit to the susceptibility data. However, the listed CEF parameters also qualitatively explain the magnetization data as shown in Fig.~\ref{Fig4}(b).    The energy eignevalues depicted in Fig.~\ref{Fig5} provide the closest fit to the  Schottky heat capacity (to be discussed later) and can be taken as the first order esimates.  From the mean field theory, one can obtain a rough estimate of the $B_2^0$ parameter which is related to the paramagnetic Weiss temperature and the exchange constant by the following relation~\cite{jensen1991rare}

\begin{equation}
k_{\rm B}\theta_a = \frac{1}{3}J(J+1)\mathcal{J}_{\rm ex}^a + \frac{2}{5}\left(J - \frac{1}{2}\right)\left(J + \frac{3}{2}\right)B_2^0,
\label{eqn6}
\end{equation}
and
\begin{equation}
k_{\rm B}\theta_c = \frac{1}{3}J(J+1)\mathcal{J}_{\rm ex}^c - \frac{4}{5}\left(J - \frac{1}{2}\right)\left(J + \frac{3}{2}\right)B_2^0.
\label{eqn7}
\end{equation}

Assuming, an isotropic two-ion interaction $\mathcal{J}_{\rm ex}^a = \mathcal{J}_{\rm ex}^c = \mathcal{J}_{\rm ex}$ and using   $J= 15/2$  and the paramagnetic Curie-Weiss temperature $\theta_{\rm p}$ values, obtained in section 3.3, the value of $B_2^0$ was estimated to be $-0.092$~K.   Given the assumptions involved in the point charge model and the limitations of the mean filed theory, this value may be considered to be in fairly good agreement with the  $B_2^0$  value obtained from the CEF analysis.  It thus provides some additional support to our final choice of crystal field parameters and the resulting crystal field levels.  

\subsection{Heat capacity studies}

%*********************FIGURE 6********************************
\begin{figure}[!]
\includegraphics[width=0.5\textwidth]{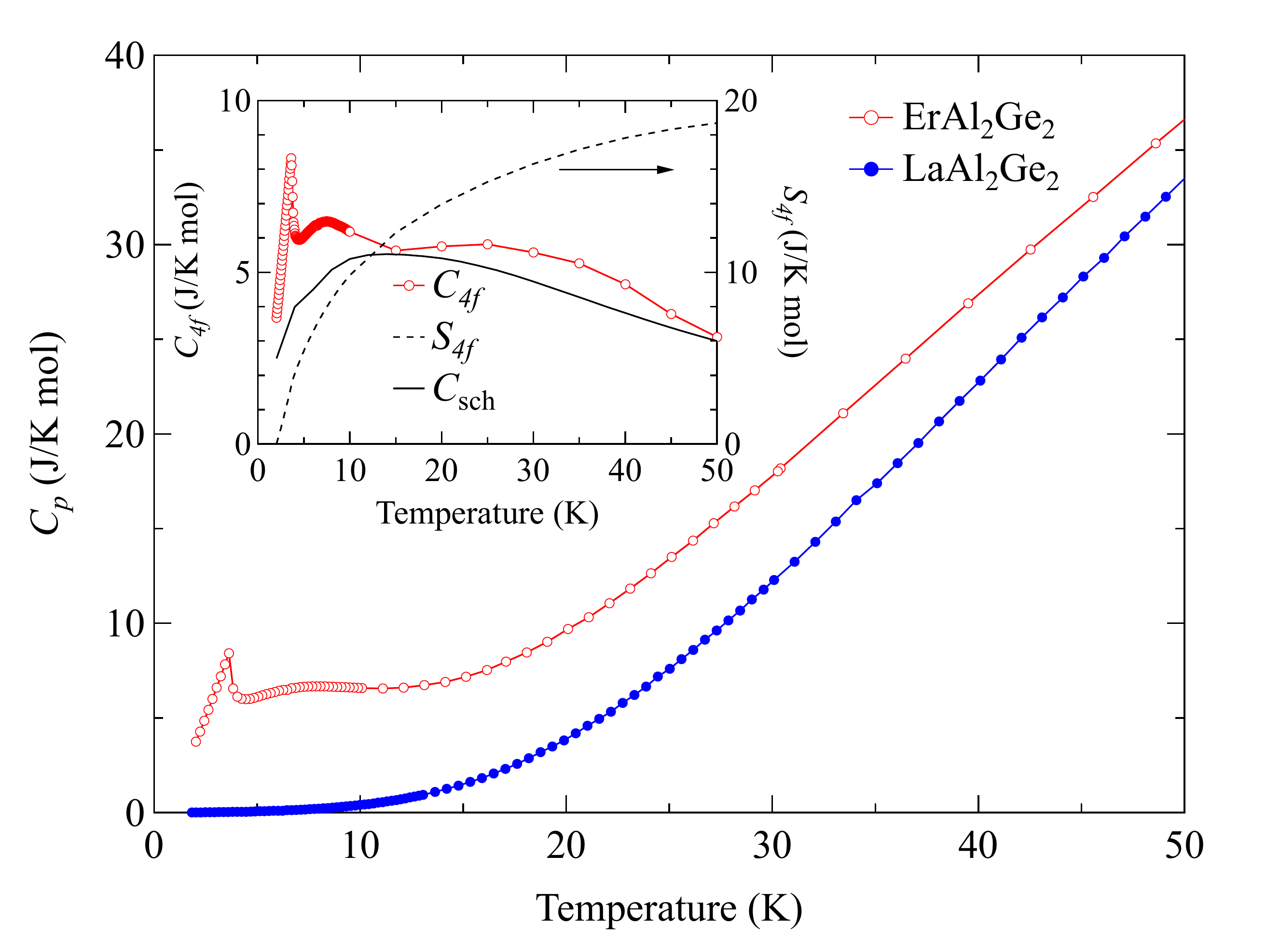}
\caption{(Color online) Temperature variation of heat capacity of ErAl$_2$Ge$_2$ and LaAl$_2$Ge$_2$. Inset shows the temperature variation of the $4f$-derived heat capacity and the corresponding entropy of ErAl$_2$Ge$_2$ and the solid line is the calculated Schottky heat capacity based on CEF calculations.}
\label{Fig6}
\end{figure}
%****************************************************************

The temperature dependence of the specific heat capacity of a single crystalline \EAG\ and its non-magnetic reference polycrystalline LaAl$_2$Ge$_2$ in the temperature range $2-50$~K is shown in the main panel of Fig.~\ref{Fig6}.  The heat capacity of the magnetic compound \EAG\ is greater than the non-magnetic compound in the entire temperature range, due to the additional $4f$-derived contribution to the heat capacity $C_{4f}$.    A very sharp $\lambda$-like transition is observed at $T_{\rm N} = 4$~K suggesting the bulk nature of the magnetic ordering in \EAG.  We have  estimated the $4f$-derived contribution to the heat capacity $C_{4f}$,  by subtracting the phononic contribution (taken to be identical to that of LaAl$_2$Ge$_2$) from  the heat capacity of \EAG\ and plotted it in the inset of Fig.~\ref{Fig6}.  Above the magnetic transition, a broad peak is observed in the magnetic part of the heat capacity, which is mainly attributed to the Schottky heat capacity arising due to the thermal population of the several low lying crystalline electric field states as the temperature is increased.  Theoretically, the Schottky heat capacity can be calculated using the energy levels obtained from the CEF fitting of the magnetic susceptibility data.  The solid curve in the inset of Fig.~\ref{Fig6}, representing the CEF derived Schottky heat capacity is in a semi-quantitative agreement with the experimentally observed values.    An estimate of the entropy change with temperature has also been carried out and it is shown in the inset of Fig.~\ref{Fig6}.  The entropy increases very rapidly and attains a value of nearly 20~J/K$\cdot$mol, thus suggesting several low lying CEF states.

\section{Summary}

In summary, we have successfully grown the single crystals of \EAG\ and studied its anisotropic physical properties.  From the susceptibility and magnetization  data, we infer that  the easy axis of magnetization lies in the $ab$-plane.  The anisotropy in the magnetic susceptibility can be qualitatively explained by our CEF analysis.  It has been shown that the 16 fold ($2J+1$) degenerate level of the free Er$^{3+}$ ion splits into 8 doublets with an overall separation of just 108~K with several levels lying at low energies.  The estimated crystal field energy levels account semi-quantitatively for the experimentally observed Schottky heat capacity.  The electrical resistivity revealed that a  superzone gap forms in \EAG\ below the antiferromagnetic transition. The exchange interaction, as reflected by the 
magnetic transition temperature in Ho and Er analogs decreases as one moves towards the higher rare-earth side, which is in  accordance with the deGennes scaling. It would be interesting to study the next compound in the series TmAl$_2$Ge$_2$ where, based on the CEF theory, one can expect to see the change of the easy axis compared to \EAG\ and that work is planned as a future study.

\section*{References}

%\bibliography{mybibfile}
\bibliography{Ref}

\end{document}